\documentclass[12pt, a4paper]{article}
\usepackage[utf8]{inputenc}
\usepackage{physics} 
\usepackage{amsmath}
\usepackage{amsfonts}
\usepackage{amssymb}
\usepackage{graphicx}
\usepackage{authblk}
\usepackage{subcaption}
\usepackage{cite}

\usepackage[left = 1.5cm, right = 1.5cm, top = 1cm, bottom = 2cm]{geometry}
\begin{document}

\begin{center}
{}
{\large {\bf Observational Constraints on a Spinor Field Generalized Chaplygin Gas Model in a Spherically Symmetric FLRW Spacetime}}

\vskip .5cm
\textbf{ Mahendra Goray$^{a}$, Bijan Saha$^{c, d}$}
\end{center}
\vskip0.1cm
$^a$Department of Physics, S.P. College, Sido Kanhu Murmu University, Dumka-814101, Jharkhand, India\\
$^b$Laboratory of Information Technologies, Joint Institute for Nuclear Research,
141980 Dubna, Moscow region, Russia\\
$^c$Peoples' Friendship University of Russia (RUDN University), 6 Miklukho-Maklaya Street, Moscow, Russian Federation\\

\textbf{E-mail:} goraymahendra92@gmail.com \textbf{(M.Goray)}, bijan@jinr.ru \textbf{(B.Saha)} \\

\abstract{
Despite the remarkable success of the standard 
$\Lambda$CDM model in describing the evolution of the universe, several unresolved issues remain, such as the true nature of dark energy, fine-tuning problems, and the persistent Hubble tension. Motivated by these shortcomings, we construct a spinor field-based Generalized Chaplygin Gas (GCG) model that unifies dark matter and dark energy within a spherically symmetric Friedmann–Lemaître–Robertson–Walker (FLRW) spacetime. This framework incorporates a nonlinear spinor field and considers an open universe geometry. We constrain the model parameters using the latest observational datasets, including Type Ia supernovae from the binned Pantheon compilation, Hubble parameter measurements from cosmic chronometers (CC) and SDSS, including baryon acoustic oscillation (BAO) data. Employing Markov Chain Monte Carlo (MCMC) sampling techniques, we obtain best-fit values that indicate the spinor GCG model provides a competitive and viable alternative to the $\Lambda$CDM, particularly in the late-time universe. Furthermore, the model predicts a lower present-day Hubble constant, offering a potential resolution to the Hubble tension. The results highlight the rich phenomenology of spinor fields and their possible role in the dynamics of dark energy through spacetime interaction.
}

\section*{I. Introduction}
Modern cosmology has achieved remarkable success in describing the large-scale structure and evolution of the universe, culminating in the vacuum energy model (or cosmological constant $\Lambda$), especially the $\Lambda$CDM model, a simple yet highly predictive framework \cite{Planck2018}. Including the anisotropies in the Cosmic Microwave Background (CMB) radiation \cite{Planck2018}, the Large-Scale Structure (LSS) of the universe \cite{BOSS2017}, and the luminosity distances of Type Ia supernovae (SN Ia) \cite{Pantheon2018}. Despite its tremendous progress, it still faces several significant challenges, both theoretical and observational. In particular, the nature of dark energy (DE) and dark matter (DM) \cite{PeeblesRatra2003, Frieman2008}, and CMB anomalies \cite{Schwarz2016}. Additionally, persistent tensions between early and late universe measurements of the Hubble parameter $H_{0}$ \cite{Riess2021, DiValentino2021}, motivate the investigation of alternative frameworks beyond $\Lambda$CDM.

To address these issues, a wide variety of alternative models have been proposed. These include dynamic dark energy models based on scalar fields (e.g., quintessence, phantom, k-essence) \cite{Caldwell1998, ArmendarizPicon2001}, as well as modified gravity theories \cite{Clifton2012}. In parallel, unified dark sector models have gained traction by aiming to describe dark matter and dark energy within a single fluid framework \cite{Bertone2005}. Among these, the Generalized Chaplygin Gas (GCG) model is characterized by the exotic equation of state \cite{Kamenshchik2001, Bento2002}, which is a mixture of DE and DM, behaves as a dust-like matter at early times, and behaves like a cosmological constant at the late stage. Reducing two unknown physical entities into a single one, the accelerated expansion of the universe arises from an uncanceled cosmological constant.

Nevertheless, the GCG model in its conventional hydrodynamical formulation faces challenges in structure formation and in matching perturbative behavior with observations \cite{Sandvik2004}. This has led to alternative realizations of the model, including representations based on fundamental fields. One such extension involves a self-interacting nonlinear spinor field \cite{Ribas2005, Saha2006}, which naturally yields an effective GCG-type equation of state when embedded in a spherically symmetric FLRW background. The spinor field framework is motivated not only by its rich phenomenological behavior but also by its theoretical foundation in field theory and its potential for coupling to gravity in a consistent manner \cite{ArmendarizPicon2003, Kremer2003}.

In this work, we investigate a cosmological model in which the generalized Chaplygin gas arises from a self-interacting spinor field \cite{Ribas2005, Saha2006}. We derive the background dynamics and construct a model for the Hubble parameter $H(z)$ using a phenomenological form inspired by the effective energy density and pressure of the spinor field \cite{Kremer2003}. The model parameters are constrained using a joint likelihood analysis involving current observational data sets: the Hubble parameter measurements from cosmic chronometers and SDSS (OHD + SDSS) \cite{Moresco2012, Sahni2014}, the binned Pantheon compilation of Type Ia supernovae (SNb) \cite{Pantheon2018}, and two low-redshift Baryon Acoustic Oscillation (BAO) measurements \cite{Beutler2011, Ross2015}. A Markov Chain Monte Carlo (MCMC) sampling is employed to extract the posterior distributions of the parameters, and several derived cosmological quantities, including the deceleration parameter $q(z)$, the effective equation of state $w(z)$, and the distance modulus $\mu(z)$. The statistical model selection criteria, such as the reduced chi-square, Akaike Information Criterion (AIC), and Bayesian Information Criterion (BIC), are used to assess the model's performance relative to $\Lambda$CDM. The analysis is carried out using the affine-invariant ensemble sampler \texttt{emcee} \cite{ForemanMackey2013}. As our model was constructed as a unified spinor field GCG model for a late-time universe, we excluded the high-redshift CMB data ($R, l_{A}, z_{*}$) from the MCMC sampling.

The structure of this paper is organized as follows: In Section II, we construct the spinor field GCG model from a mathematical background, using a spherically symmetric FLRW space-time. Section III is dedicated to the observational data sets and the MCMC methodology used for parameter estimation. In Section IV, we analyze the numerical results obtained from the simulations. Finally, the main conclusions of the study are summarized in Section V.

\section*{II. The spinor field GCG model}

Since the universe is practically isotropic and homogeneous, we consider the spherically symmetric FLRW line element \cite{Narlikar}:
\begin{equation}
ds^2 = dt^2 - a^2(t)\left[\frac{dr^2}{1 - k r^2} + r^2 d\theta^2
+ r^2 \sin^2{\theta} \, d\phi^2\right], \label{frwsph}
\end{equation}
where $a(t)$ is the cosmological scale factor and $k$ is the curvature index of the spatial geometry, indicating a flat, open, or closed universe for $k = 0$, $k = -1$, and $k = +1$, respectively. In this study, motivated by observational evidence, we consider an open universe with $k = -1$.

We construct the action such that the spinor field is minimally coupled to gravity. The action is given by 
\begin{eqnarray}
{\cal S} = \int \sqrt{-g} \left[\frac{R}{2 \kappa} + L_{\rm sp}
\right] d\Omega, \label{action}
\end{eqnarray}
where $\kappa = 8 \pi G$ is Einstein’s gravitational constant, $R$ is the scalar curvature, and $L_{\rm sp}$ is the Lagrangian density of the spinor field, expressed as \cite{SahaPRD2001}
\begin{eqnarray}
L_{\rm sp} = \frac{\imath}{2} \left[\bar{\psi} \gamma^{\mu} \nabla_{\mu}
\psi - \nabla_{\mu} \bar{\psi} \, \gamma^{\mu} \psi \right] - m_{\rm sp} \bar{\psi} \psi
- \lambda F(K). \label{lspin}
\end{eqnarray}

Here, $m_{\rm sp}$ and $\lambda$ denote the spinor mass and the self-coupling constant, respectively. The function $F(K)$ represents a nonlinear term introduced to preserve Lorentz invariance in the spinor field equations \cite{SahaPRD2001}. The corresponding spinor field equations are given by
\begin{subequations}
\label{speq}
\begin{align}
\imath\gamma^\mu \nabla_\mu \psi - m_{\rm sp} \psi - \mathcal{D} \psi - \imath \mathcal{G} \gamma^5 \psi &= 0, \label{speq1} \\
\imath \nabla_\mu \bar{\psi} \gamma^\mu +  m_{\rm sp} \bar{\psi} + \mathcal{D} \bar{\psi} + \imath \mathcal{G} \bar{\psi} \gamma^5 &= 0, \label{speq2}
\end{align}
\end{subequations}
where
\[
\mathcal{D} = 2 \lambda F_K b_1 S, \quad \mathcal{G} = 2 \lambda F_K b_2 P, \quad F_K = \frac{dF}{dK}.
\]
In view of the spinor field equations \eqref{speq}, the spinor field Lagrangian \eqref{lspin} can be rewritten as
\begin{align}
L = \lambda \left(2 K F_K - F\right). \label{lags1}
\end{align}

The components of the Einstein field equations for the spinor field in the FLRW metric read
\begin{subequations}
\label{EMTc}
\begin{align}
T_0^{\;0} &=  m_{\rm sp} S + \lambda F \;=: \;\varepsilon, \label{00f} \\
T_1^{\;1} &=  T_2^{\;2} = T_3^{\;3} = -\lambda\!\left(2 K F_K-F\right) \;=: -p, \label{iif}\\
T^{1}_{\;3} &= \frac{a \cos\theta}{4\sqrt{1 - k r^2}}\,A^{0}, \label{13f}\\
T^{0}_{\;1} &= \frac{\cot\theta}{4 r \sqrt{1 - k r^2}}\,A^{3}, \label{10f}\\
T^{0}_{\;2} &= -\frac{3}{4}\sqrt{1 - k r^2}\,A^{3}, \label{02f}\\
T^{0}_{\;3} &=  \frac{3}{4}\sqrt{1 - k r^2}\,\sin\theta\,A^{2}
              -\frac{1}{2}\cos\theta\,A^{1}. \label{03f}
\end{align}
\end{subequations}
Here \(p\) and \(\varepsilon\) denote the pressure and energy density, respectively \cite{bsaha2025}.  
Because the off-diagonal components of the energy–momentum tensor vanish identically, equations \eqref{13f} - \eqref{03f} impose the following restrictions on the spinor field:
\begin{align}
    A^0 &= 0, \quad A^3 = 0, \quad A^1 = \frac{3}{2}\sqrt{1-kr^2}\, \tan\theta A^2. \label{res} 
    \end{align}
The Einstein field equations reduce to
\begin{subequations}
\label{ein2}
\begin{align}
2\frac{\ddot a}{a} + \left(\frac{\dot a^{\,2}}{a^{2}} + \frac{k}{a^{2}}\right)
    &= -8\pi G\,p, \label{ein11m}\\[4pt]
3\left(\frac{\dot a^{\,2}}{a^{2}} + \frac{k}{a^{2}}\right)
    &=  8\pi G\,\varepsilon. \label{ein00m}
\end{align}
\end{subequations}
Using Eq.~\eqref{ein00m}, Eq.~\eqref{ein11m} can be recast as
\begin{equation}
\frac{\ddot a}{a} = -\frac{4\pi G}{3}\bigl(\varepsilon + 3p\bigr). \label{accel}
\end{equation}
Combining Eqs.~\eqref{00f}, \eqref{iif}, and \eqref{accel} yields
\begin{equation}
\ddot a = -\frac{4\pi G}{3}\!\left(m_{\rm sp} S - 2\lambda F + 6\lambda K F_K\right)a. \label{acceln}
\end{equation}
Accordingly,
\begin{align}
\dot a = \pm\sqrt{\frac{8\pi G}{3}\,\varepsilon\,a^{2} - k}
       = \pm\sqrt{\frac{8\pi G}{3}\bigl(m_{\rm sp} S + \lambda F\bigr)a^{2} - k}. \label{vel}
\end{align}


Starting from Eqs.~\eqref{ein2}, we can recast the dynamics in the form
\begin{subequations}
\label{ein2a}
\begin{align}
\dot a &= H\,a, \label{HCsys}\\
\dot H &= -\frac{3}{2}\,H^{2} - \frac{1}{2}\,\frac{k}{a^{2}}
         - 4\pi G\lambda\bigl(2K F_{K}-F\bigr), \label{ein11msys}\\
H^{2} &= \frac{8\pi G}{3}\bigl(m_{\rm sp}S+\lambda F\bigr)
        - \frac{k}{a^{2}}, \label{ein00msys}
\end{align}
\end{subequations}
where \(H\) is the Hubble parameter.  As noted earlier, we adopt an open universe (\(k=-1\)) and, for the dark-energy sector, assume a massless spinor field (\(m_{\rm sp}=0\)).  Under these conditions Eqs.~\eqref{ein2a} become, after rewriting in terms of the redshift \(z=a_{0}/a-1\) \(\bigl(a_{0}=1\bigr)\),
\begin{subequations}
\label{ein2az}
\begin{align}
\frac{da}{dz} &= -\frac{a_{0}}{(1+z)^{2}}, \label{HCsysz}\\
\frac{dH}{dz} &= \frac{1}{1+z}\!\left[H + \frac{\kappa}{6H}\bigl(\varepsilon+3p\bigr)\right]. \label{ein11msysz}
\end{align}
\end{subequations}
The conservation equation now reads
\begin{align}
\frac{d\varepsilon}{dz} &= \frac{3}{1+z}\bigl(\varepsilon+p\bigr). \label{Conservsysz}
\end{align}
The invariant $K$ in terms of red-shift reads
\[
K=\frac{K_{0}}{a^{6}} = K_{0}(1+z)^{6}, \qquad K_{0}=\text{const.}
\]

To integrate Eqs.~\eqref{ein2az} numerically, we must specify the nonlinear spinor function \(F(K)\).  Because a nonlinear spinor field can effectively mimic a wide range of cosmological fluids—including perfect fluids and various dark energy candidates such as quintessence, Chaplygin gas, and modified Chaplygin gas- we concentrate on the generalized Chaplygin gas (GCG) equation of state as a specific realization within this framework.

\begin{equation}
p = -\frac{A}{\varepsilon^{\alpha}}, \label{mchap}
\end{equation}
with positive constant\(A>0\), and $0\leq \alpha \leq 1$.  
Setting \(\varepsilon=T_{0}^{\;0}\) and \(p=-T_{1}^{\;1}\) from Eqs.~\eqref{EMTc} in Eq.~\eqref{mchap} yields
\begin{equation}
F(K)=\left(A + \lambda_{1}\,K^{(1+\alpha)/2}\right)^{1/(1+\alpha)}, \label{modchap0F}
\end{equation}
where \(\lambda_{1}\) is an integration constant.  Consequently, for a massless spinor field, we have
\begin{subequations}
\label{mchap_eps_p}
\begin{align}
\varepsilon &= \lambda \left(A+\lambda_{1}\,(1+z)^{3(1+\alpha)}\right)^{1/(1+\alpha)}, \label{mchapsped}\\
p &= -A\lambda \left(A+\lambda_{1}\,(1+z)^{3(1+\alpha)}\right)^{-\alpha/(1+\alpha)}. \label{modchapp}
\end{align}
\end{subequations}

The deceleration parameter
\(
q(z)\equiv -\tfrac{\ddot a}{a}\tfrac{1}{H^{2}}
\)
follows from Eqs.~\eqref{accel}, \eqref{ein00msys}, \eqref{00f}, and the redshift definition, giving for an open universe (\(k=-1\))
\begin{equation}
q(z)=\frac{\bigl(\varepsilon+3p\bigr)/3}{\varepsilon/3+(1+z)^{2}}. \label{Decpar}
\end{equation}
Accelerated expansion (\(q<0\)) occurs whenever \(\varepsilon+3p<0\), i.e.\ when the spinor field behaves effectively as dark energy.

\begin{figure}[!ht]
\centering
\includegraphics[width=0.8\textwidth]{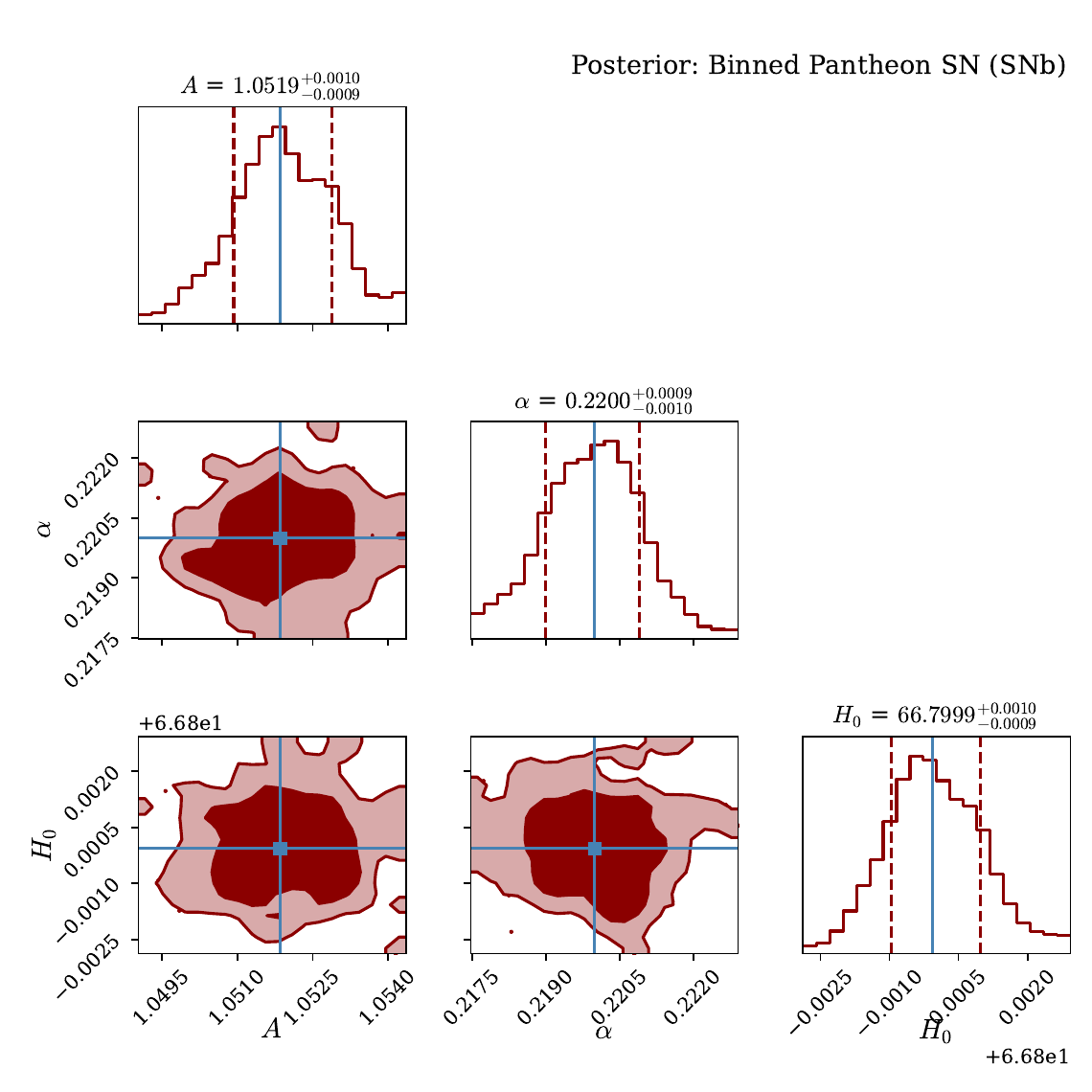}
\caption{Marginalized posterior distributions of the model parameters for the binned Pantheon data set only (SNb).}
\label{fig:SNb only}
\end{figure}

\begin{figure}[!ht]
\centering
\includegraphics[width=0.8\textwidth]{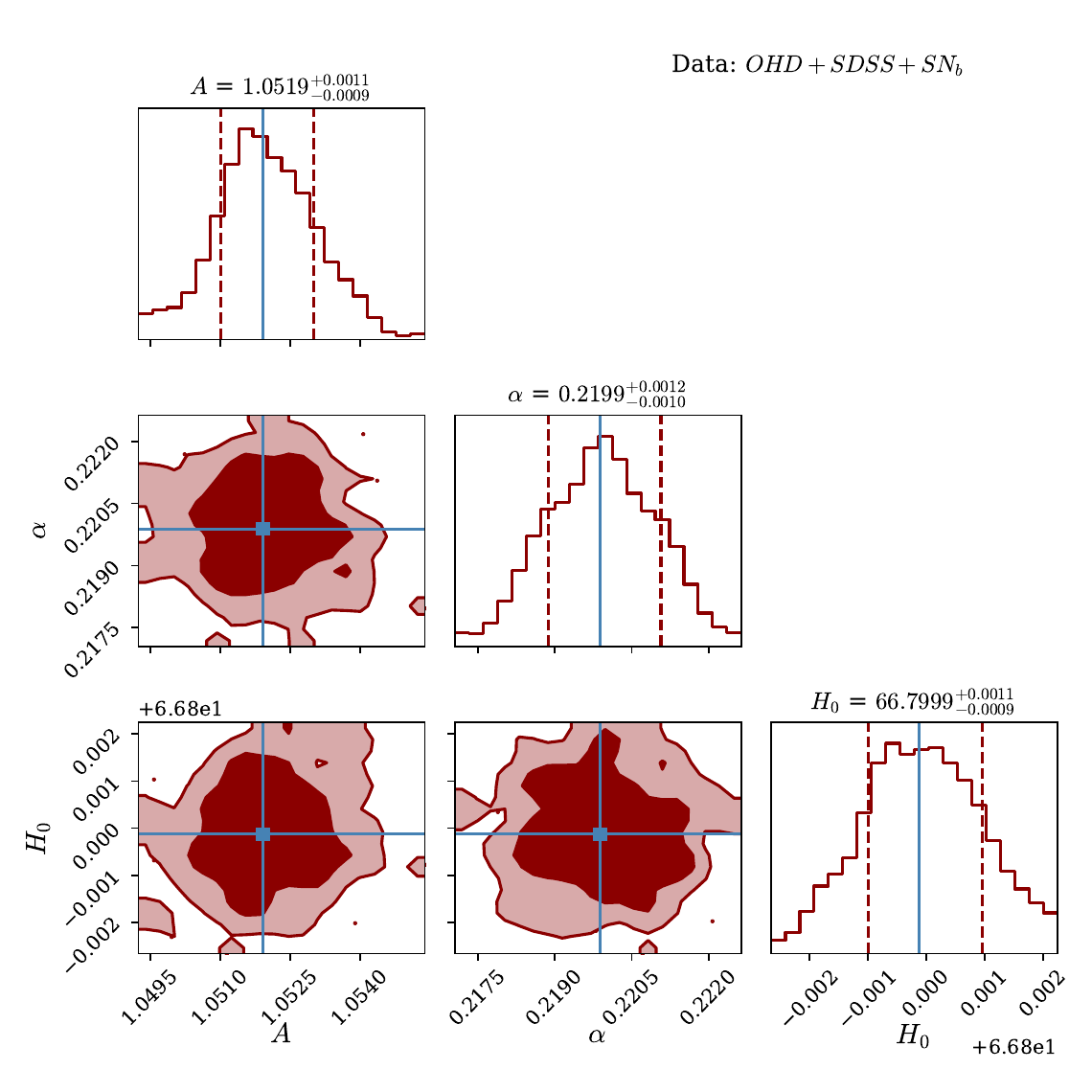}
\caption{Marginalized posterior distributions of the Spinor-GCG model parameters for Hubble data sets ($OHD + SDSS +SNb$).}
\label{fig:OHD + SDSS +SNb}
\end{figure}

\begin{figure}[!ht]
\centering
\includegraphics[width=0.8\textwidth]{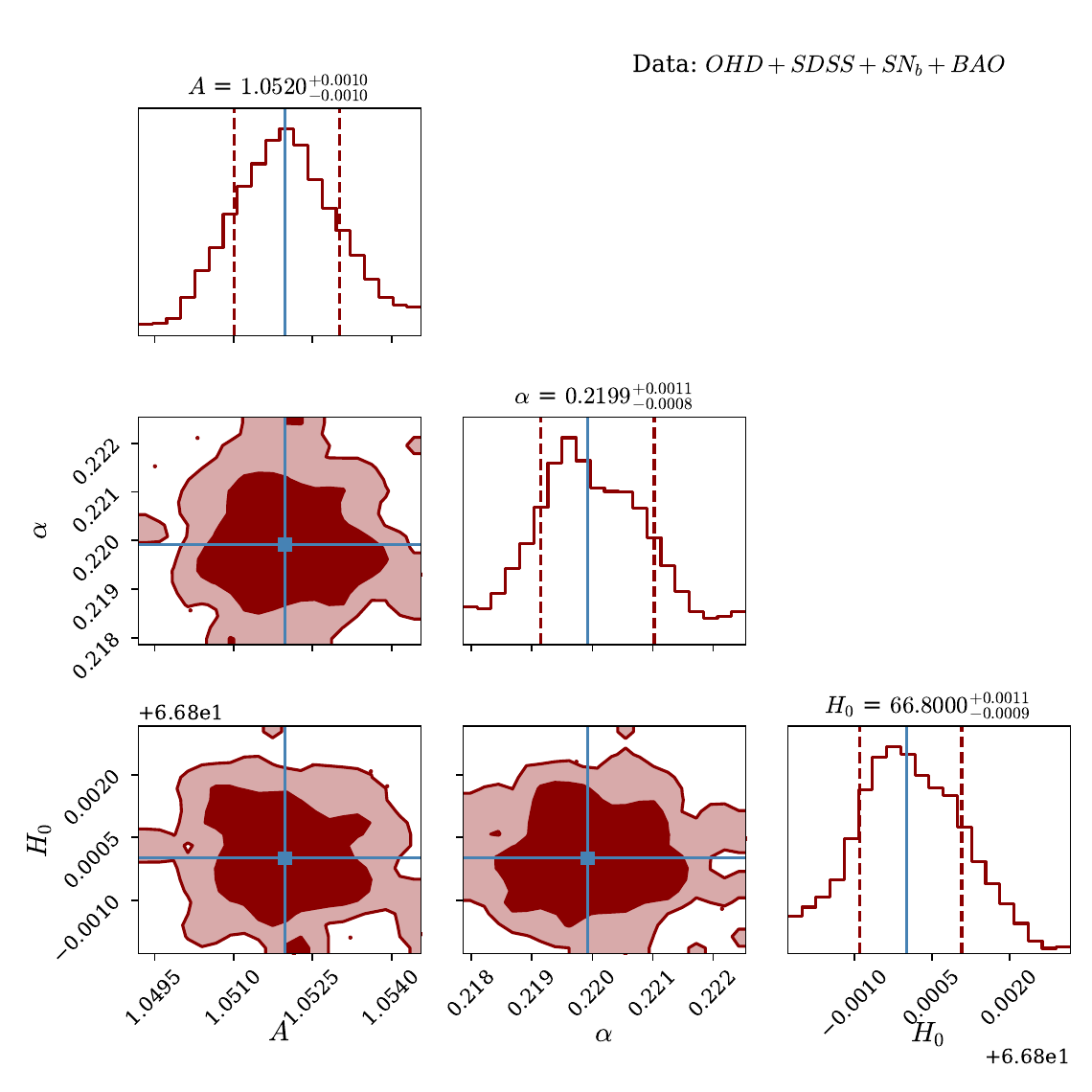}
\caption{Marginalized posterior distributions of the model parameters for the combined data sets ($OHD+SDSS+SNb+BAO$).}
\label{fig:OHD + SDSS +SNb+BAO}
\end{figure}

\section*{III. Observational Data and Methodology}
To constrain the free parameters of the spinor field GCG model, we utilize a combination of recent cosmological observations:

\begin{itemize}
\item \textbf{Cosmic Chronometers (CC)}: 43 measurements of the Hubble parameter $H(z)$, obtained via the differential age method, spanning the redshift interval \( z \in [0.070, 2.36] \) \cite{Moresco2012, Moresco2016}.

\item \textbf{Sloan Digital Sky Survey (SDSS)}: 48 \( H(z) \) data points extracted from BAO peak positions in galaxy clustering surveys, covering \( z \in [0.0, 2.5] \) \cite{Alam2021, Ross2015, Bautista2021}.

\item \textbf{Binned Pantheon Type Ia Supernovae (SNb)}: we use 40 binned distance modulus data from the Pantheon compilation, distributed across \( z \in [0.014, 1.612] \). To avoid local inhomogeneity effects, we exclude the first three very low-redshift points with \( z < 0.03 \) from our analysis \cite{Pantheon2018}. The observed quantity is the distance modulus \( \mu(z) \), which is related to the luminosity distance \( D_L(z) \) via
\begin{equation}
\mu(z) = 5 \log_{10} \left( \frac{D_L(z)}{\text{Mpc}} \right) + 25
\end{equation}
where \( D_L(z) \) is defined in terms of the comoving distance \( D_C(z) \) as
\begin{equation}
D_L(z) = (1+z) D_C(z)
\end{equation}
with the comoving distance given by
\begin{equation}
D_C(z) = c \int_0^z \frac{dz'}{H(z')}
\end{equation}
Here, \( c \) is the speed of light in km/s, and \( H(z) \) is the Hubble parameter. The total uncertainty in the observed distance modulus includes both the statistical uncertainty from measurements and an intrinsic scatter \( \sigma_{\text{int}} \), such that the total error is
\begin{equation}
\sigma_{\mu,\text{tot}} = \sqrt{ \sigma_{\mu}^2 + \sigma_{\text{int}}^2 }.
\end{equation}
In our analysis, we adopt a constant intrinsic dispersion value of \( \sigma_{\text{int}} = 0.13 \) mag, consistent with recent literature \cite{Pantheon2018}.

\item \textbf{Baryon Acoustic Oscillation (BAO)}: two low-redshift BAO measurements, the first is from the 6dF Galaxy Survey at $z=0.106$, given as $r_{s}/D_{V}(0.106)=0.336 \pm 0.015$ \cite{6dFBAO2011}, and the second is from the SDSS-MGS survey at $z=0.15$, given as $D_{V}(0.15)/r_{s}= 4.466 \pm 0.168$ \cite{SDSSMGS2015}. We define the volume-averaged distance as 
\begin{equation}
D_V(z) = \left[ (1+z)^2 D_A^2(z) \cdot \frac{cz}{H(z)} \right]^{1/3}
\end{equation}
where $D_V(z)$ is the angular diameter distance computed from the comoving distance.

\end{itemize}

For the estimation of cosmological parameters, we employ the Markov Chain Monte Carlo (MCMC) technique as implemented in the \texttt{emcee} Python library, executed within a \textit{Jupyter}  environment using Python 3 (\textit{ipykernel}). The sampling is performed under three distinct observational configurations: (i) using only the binned Pantheon Type Ia Supernovae distance modulus data (SNb), (ii) combining the Supernovae data with Hubble parameter measurements from cosmic chronometers and SDSS (\( \mathrm{OHD} + \mathrm{SDSS} + \mathrm{SNb} \)), and (iii) a comprehensive combination of all available data including BAO measurements (\( \mathrm{OHD} + \mathrm{SDSS} + \mathrm{SNb} + \mathrm{BAO} \)). Throughout all MCMC runs, we fix the spinor field coupling constant to \( \lambda = 1.75 \) and the integration constant to \( \lambda_1 = 0.332 \), ensuring both numerical stability and physical consistency of the model evolution.

 We explore a three-dimensional parameter space \( \{A, \alpha , H_0\} \). Here, $A$, $\alpha$ represent a positive constant and \( H_0 \) denotes the present-day Hubble parameter. The prior ranges adopted for the parameters are: 
 $0 < A < 3, \quad 0 < \alpha < 1.5, \quad 50 < H_0 < 90$.
The MCMC simulations were run using 200 walkers and 5000 steps. For comparison, we also perform MCMC sampling for the standard $\Lambda$CDM model using the same combinations of observational datasets. The parameter space explored in this case is two-dimensional, consisting of $\{\Omega_{m}, H_{0}\}$, which allows for a direct comparison with the spinor field GCG model.

The total chi-square function employed in the likelihood analysis is defined as
\begin{equation}
\chi^2_{\text{total}} = \chi^2_{\mathrm{OHD}} + \chi^2_{\mathrm{SDSS}} + \chi^2_{\mathrm{SNb}} + \chi^2_{\mathrm{BAO}},
\end{equation}
where \( \chi^2_{\mathrm{OHD}} \), \( \chi^2_{\mathrm{SDSS}} \), \( \chi^2_{\mathrm{SNb}} \), and \( \chi^2_{\mathrm{BAO}} \) represent the individual contributions from the cosmic chronometer (OHD), Sloan Digital Sky Survey (SDSS), Pantheon binned Type Ia Supernovae (SNb), and baryon acoustic oscillation (BAO) datasets, respectively. Each chi-square term is computed via
\begin{equation}
\chi^2 = \sum_{i=1}^{N} \left( \frac{O_i - T_i}{\sigma_i} \right)^2,
\end{equation}
where \( O_i \) denotes the observed value, \( T_i \) is the corresponding theoretical prediction from the model, \( \sigma_i \) is the measurement uncertainty, and \( N \) is the total number of data points in the respective dataset \cite{MCMC2002, Pantheon2018}.

\section*{IV. Results}
To quantify the constraints on the spinor field Generalized Chaplygin Gas (GCG) model parameters, we present the two-dimensional marginalized posterior distributions and one-dimensional likelihoods in the form of corner plots. The MCMC contours correspond to the datasets: (i) SNb-only (Figure~\ref{fig:SNb only}), (ii) OHD + SDSS + SNb (Figure~\ref{fig:OHD + SDSS +SNb}), and (iii) OHD + SDSS + SNb + BAO (Figure~\ref{fig:OHD + SDSS +SNb+BAO}). The model parameter space includes the GCG coefficient \( A \), the Chaplygin index \( \alpha \), and the Hubble constant \( H_0 \). The MCMC results demonstrate tight constraints with nearly symmetric Gaussian posteriors centered around \( A \approx 1.052 \), \( \alpha \approx 0.220 \), and \( H_0 \approx 66.8~\mathrm{km~s^{-1}~Mpc^{-1}} \). The contours show high consistency between different datasets, with progressively improved precision as additional observational data (BAO) are included. This robustness across combinations supports the viability of the spinor field GCG scenario as an effective cosmological model in the late-time universe. These MCMC results are summarized in Table~1.

\begin{figure}[!ht]
  \centering
  \includegraphics[width=0.85\textwidth]{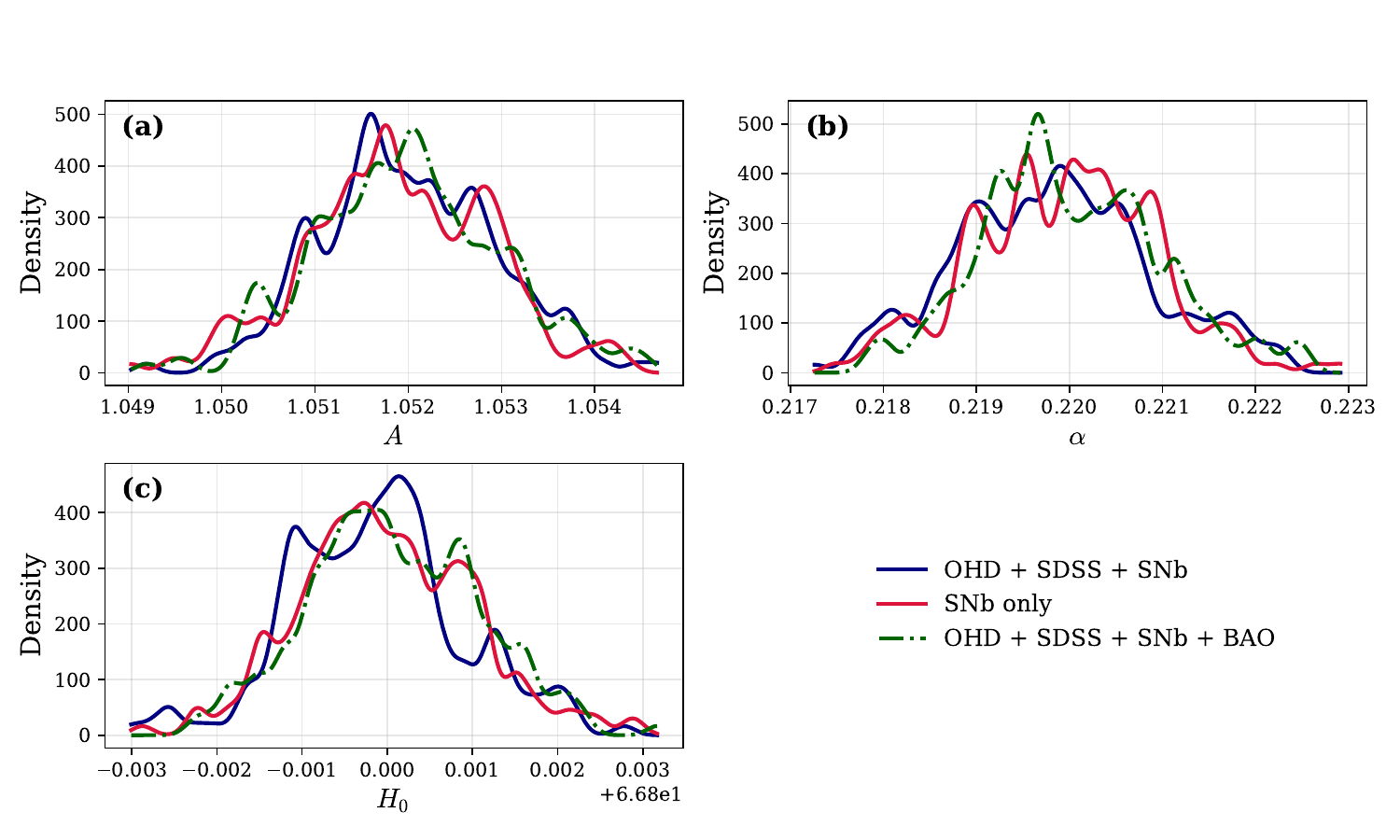}
  \caption{Marginal posterior distributions of the model parameters using different datasets. Panel (a): constant parameter $A$, (b): parameter $\alpha$, (c): Hubble constant $H_0$.}
  \label{fig:kde_comparison}
\end{figure}

\begin{table}[htbp]
\centering
\caption{Best-fit values of the model parameters using different datasets. Uncertainties correspond to the 68\% confidence interval.}
\vspace{0.2 cm}
\label{tab:best_fit_params}
\begin{tabular}{lccc}
\hline
\vspace{0.1 cm}
\textbf{Dataset} & \(\boldsymbol{A}\) & \(\boldsymbol{\alpha}\) & \(\boldsymbol{H_0}\) [km/s/Mpc] \\
\hline
\vspace{0.1 cm}
\textbf{Binned Pantheon SN (SNb)}  &
\(1.0519^{+0.0010}_{-0.0009}\) &
\(0.2200^{+0.0009}_{-0.0010}\) &
\(66.7999^{+0.0010}_{-0.0009}\) \\
\vspace{0.1 cm}
\textbf{OHD + SDSS + SNb} &
\(1.0519^{+0.0011}_{-0.0009}\) &
\(0.2199^{+0.0012}_{-0.0010}\) &
\(66.7999^{+0.0011}_{-0.0009}\) \\
\textbf{OHD + SDSS + SNb + BAO} &
\(1.0520^{+0.0010}_{-0.0010}\) &
\(0.2199^{+0.0011}_{-0.0008}\) &
\(66.8000^{+0.0011}_{-0.0009}\) \\
\hline
\end{tabular}
\end{table}

Table~\ref{tab:best_fit_params} presents the best-fit values of the model parameters \(A\), \(\alpha\), and \(H_0\) derived from different combinations of observational datasets using MCMC analysis. The quoted uncertainties correspond to the 68\% confidence intervals obtained from the marginalized posterior distributions. These model parameters are displayed in Figure~\ref{fig:kde_comparison}, the one-dimensional marginalized posterior distributions as inferred from three different combinations of datasets. The smooth Kernel Density Estimates (KDE) clearly illustrate how the inclusion of additional data progressively narrows the posterior distributions, particularly for the parameters \( A \) and \( \alpha \). This tightening of the contours reflects improved constraints on the model parameters due to complementary information from Hubble and BAO data. The consistency between the peaks across all three curves also supports the robustness of the inferred parameter values. Notably, the Hubble constant \( H_0 \) is well-constrained, with the full dataset combination yielding the tightest posterior width, demonstrating the model's capability to accommodate current observational data.

\begin{figure}[!ht]
\centering
\includegraphics[width=0.85\textwidth]{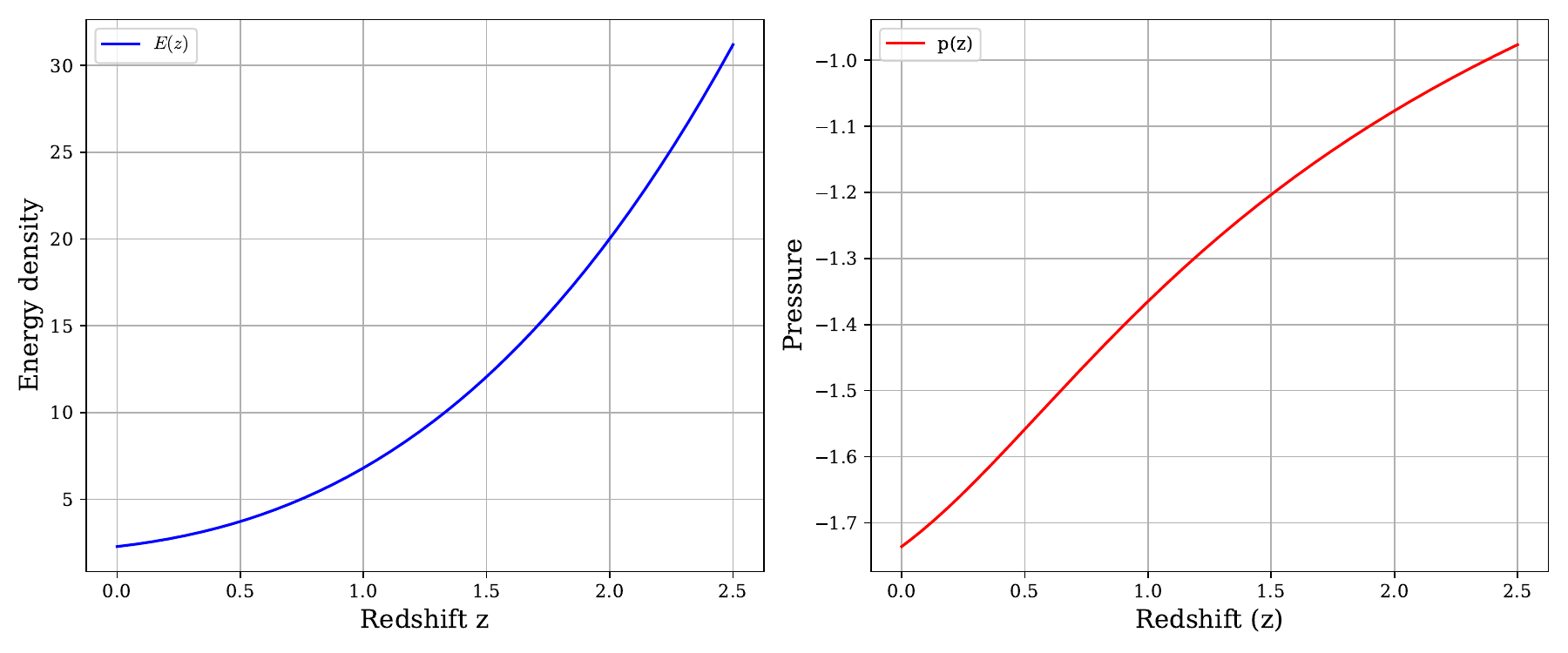}
\caption{Redshift evolution of the energy density \( \varepsilon(z) \) (left panel) and pressure \( p(z) \) (right panel) for the spinor field Generalized Chaplygin Gas (GCG) model using the best-fit parameters.}
\label{fig:Energy Density}
\end{figure}

In Figure~\ref{fig:Energy Density}, the function \( \varepsilon(z) \) represents the total energy density of the effective cosmic fluid, while \( p(z) \) captures the corresponding pressure. The plots illustrate the transition from a matter-dominated era to a dark-energy-like regime, reflecting the unifying behavior of the GCG model across cosmological epochs. As the universe expands, \( \varepsilon(z) \) gradually deviates from the matter-like scaling and asymptotically approaches a constant value at low redshifts (\( z \to 0 \)), and the evolution of the pressure \( p(z) \) reveals that it becomes progressively less negative (i.e., approaches zero) at higher redshifts. This behavior is characteristic of unified dark energy models such as the Generalized Chaplygin Gas (GCG), wherein the cosmic fluid transitions from a dark energy-like state at late times (with negative pressure) to a matter-like regime at early times (with negligible pressure). This interpolation supports the role of the GCG in modeling both the accelerated expansion and the matter-dominated era within a single framework.

\begin{figure}[!ht]
\centering
\includegraphics[width=0.8\textwidth]{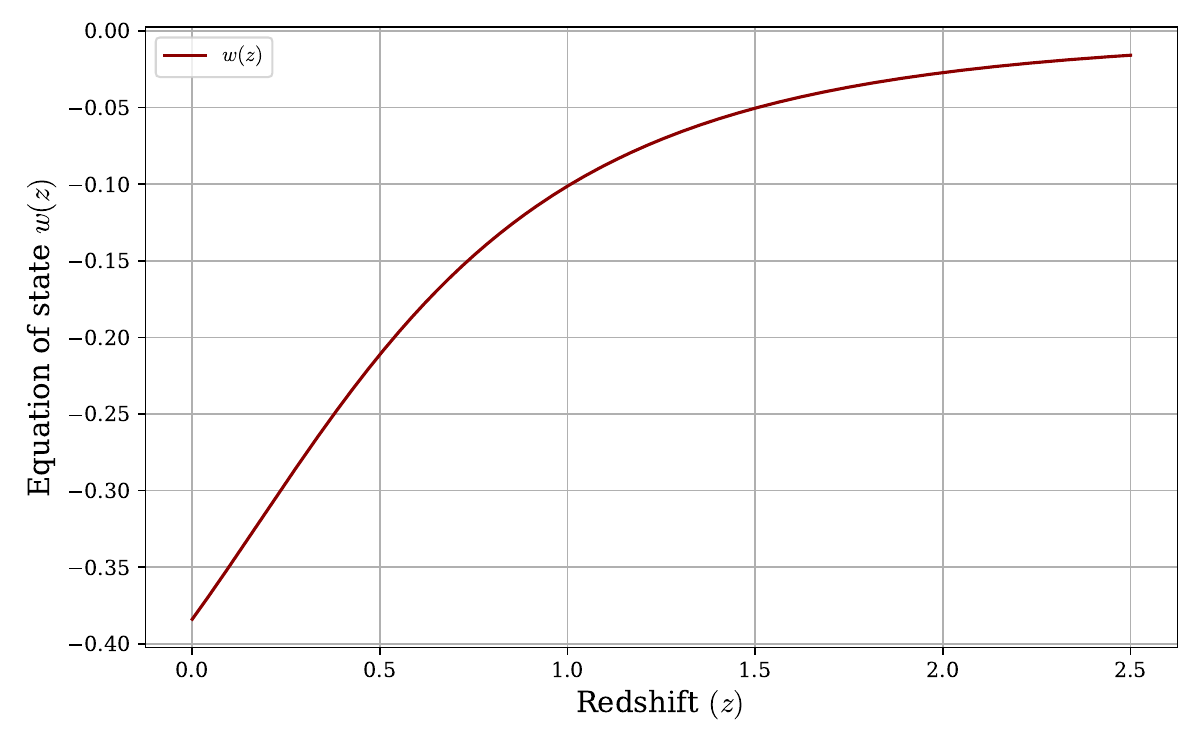}
\caption{Evolution of the equation of state $w(z)$ for the spinor field GCG model.}
\label{fig:EoS}
\end{figure}

The redshift evolution of the equation of state (EoS) parameter, as illustrated in Figure~\ref{fig:EoS}, highlights the dynamical nature of the spinor field GCG model. At the present epoch (\( z \approx 0 \)), the EoS attains a value of \( w(0) = -0.3840 \), indicating a negative-pressure component that drives cosmic acceleration. Toward the early universe (\( z = 2.5 \)), the EoS increases to \( w(2.5) = -0.0158 \), suggesting a gradual transition to a matter-like behavior. This smooth interpolation ($z=2.5 	\rightarrow z=0$)from a DM dominated phase to a DE regime reflects the unifying character of the model, in which a single spinor field fluid can effectively mimic both components.

\begin{figure}[!ht]
\centering
\includegraphics[width=0.8\textwidth]{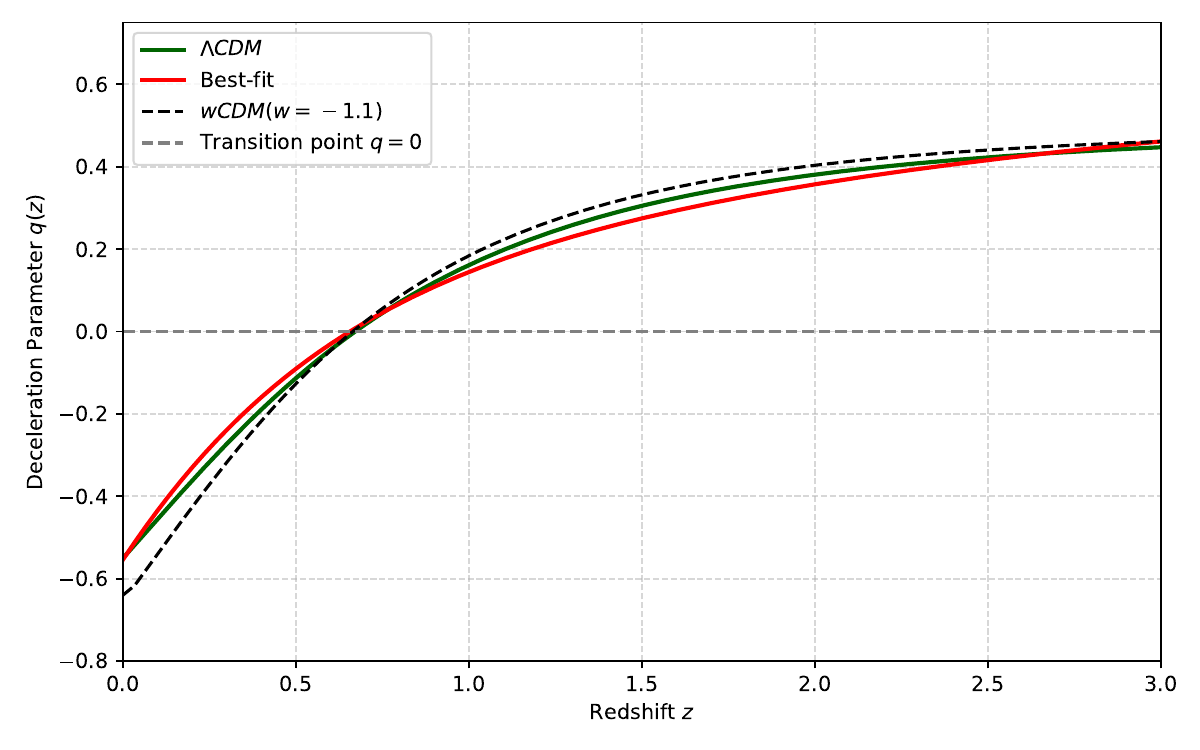}
\caption{Redshift evolution of the deceleration parameter $q(z)$ for the best-fit spinor field GCG model, compared with the $\Lambda$CDM and $w$CDM models.}
\label{fig:Deceleration}
\end{figure}

\begin{figure}[!ht]
\centering
\includegraphics[width=0.85 \textwidth]{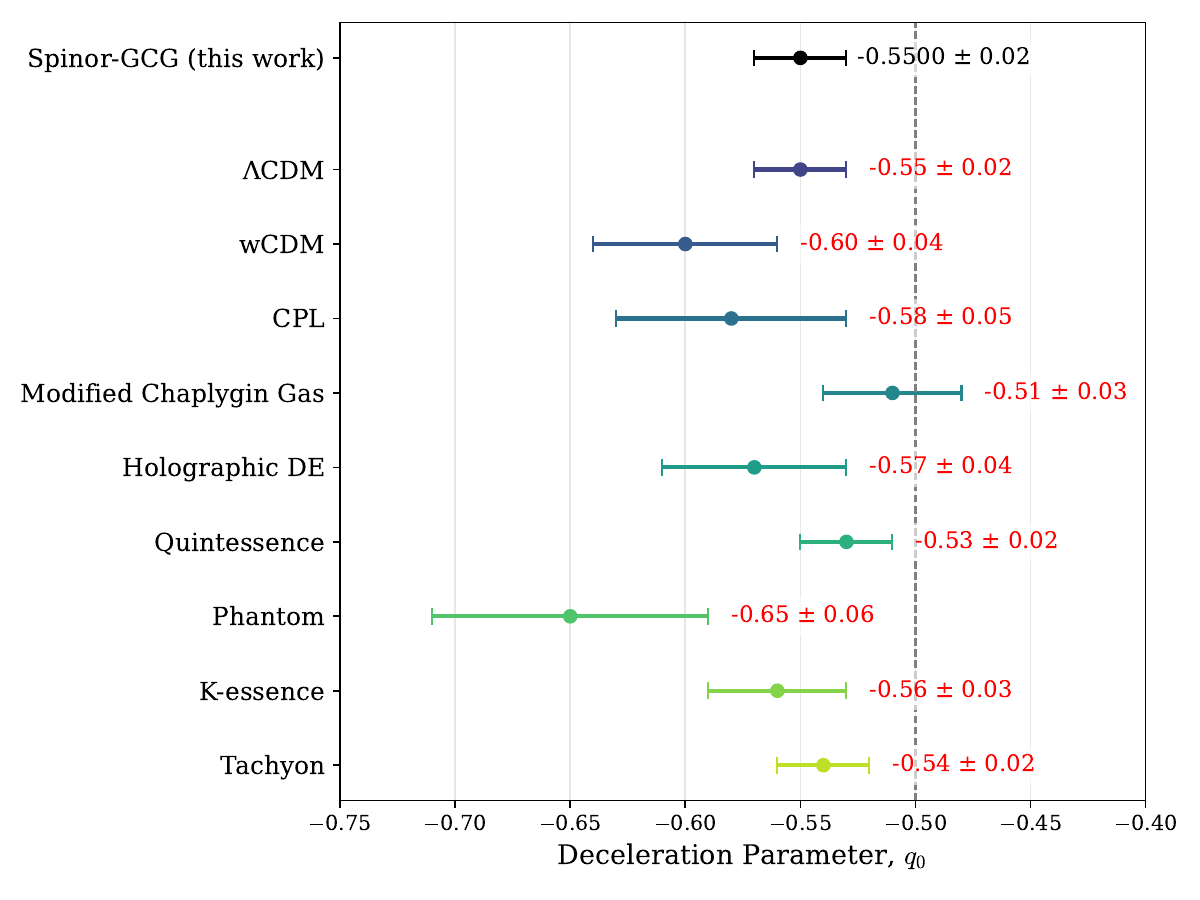}
\caption{Present deceleration \( q(0) \) for various models \cite{Planck2018, Bento2002, Lewis2002_CPL_q0, VargasDosSantos2015_q0, Santos2016_CPL}.}
\label{fig:Deceleration_comparison}
\end{figure}

To better understand the evolution of the accelerated expansion of the universe, the deceleration parameter \( q(z) \) is presented as a function of redshift in Figure~\ref{fig:Deceleration}. The best-fit result for the spinor field GCG model shows a clear transition from a decelerating to an accelerating universe. At high redshift (\( z > 1 \)), the positive value of \( q(z) \) indicates a decelerated expansion consistent with a matter-dominated era. As the redshift decreases, \( q(z) \) smoothly declines and crosses the threshold \( q=0 \) near \( z_{\text{tr}} \approx 0.67 \), marking the onset of cosmic acceleration. At the present epoch (\( z = 0 \)), we obtain a best-fit value \( q(0) = -0.5534 \), reflecting the accelerated expansion driven by the GCG component. Compared to the standard \(\Lambda\)CDM model and $w$CDM models, which yields \( q(0) \approx -0.55 \) and \( q(0) \approx -0.60 \). This close match supports the spinor field GCG model as a viable alternative to \(\Lambda\)CDM in describing the late-time cosmic acceleration. The present \( q(0) \) value for various models can also be inferred in Figure~\ref{fig:Deceleration_comparison}: for CPL \(q_0 \approx -0.58 \pm 0.05\)~\cite{Lewis2002_CPL_q0}, for \(w\)CDM \(q_0 \approx -0.60 \pm 0.04\)~\cite{VargasDosSantos2015_q0}, for the Modified Chaplygin Gas (MCG) \(q_0 \approx -0.51 \pm 0.03\)~\cite{Bento2002}, and for the spinor‑GCG model \(q_0 = -0.5500 \pm 0.020\) (this work).

\begin{figure}[!ht]
\centering
\includegraphics[width=0.8\textwidth]{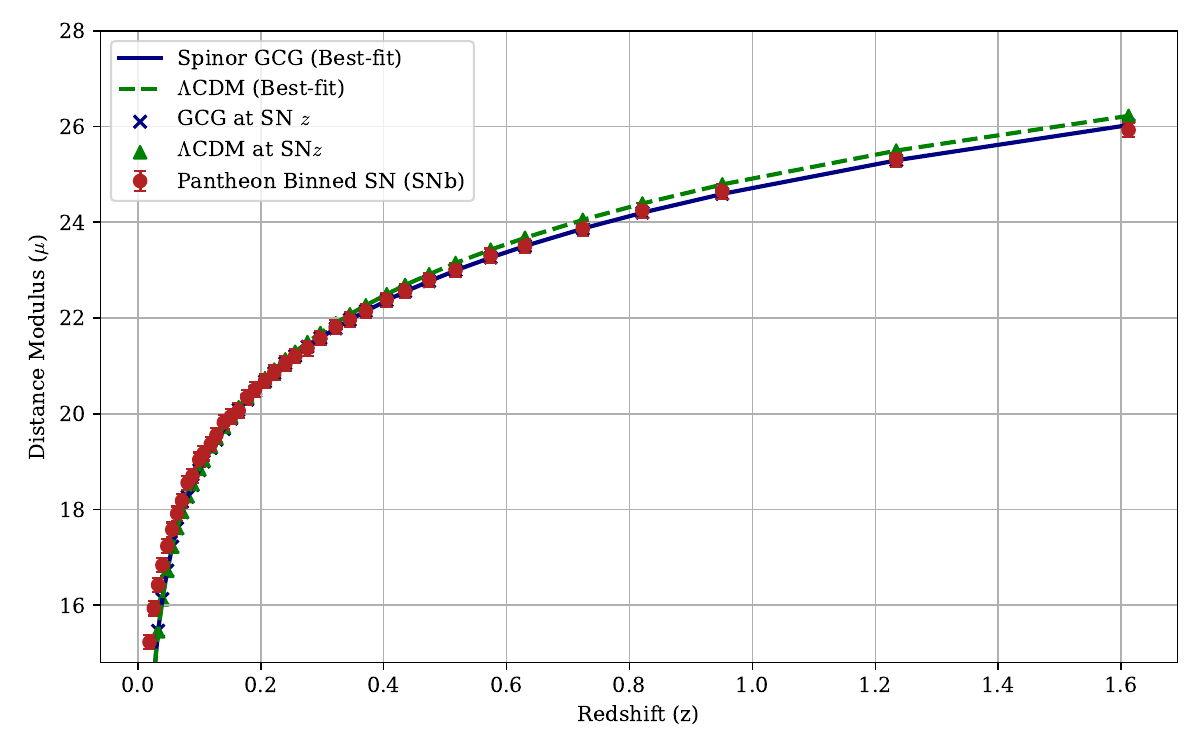}
\caption{Comparison of the best-fit theoretical distance modulus \(\mu(z)\) curves from the spinor field GCG model and the standard \(\Lambda\)CDM model with the Pantheon binned SN (SNb) data.}
\label{fig:mu_comparison}
\end{figure}

\begin{figure}[!ht]
\centering
\includegraphics[width=0.8\textwidth]{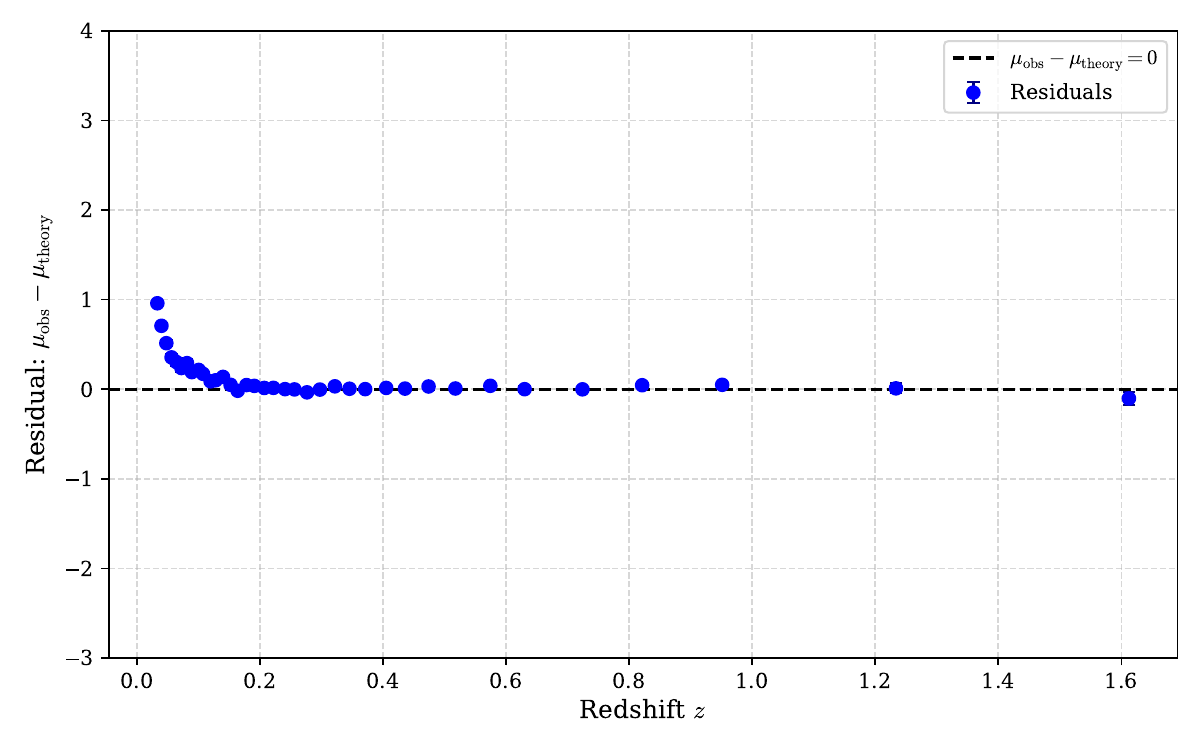}
\caption{Residuals between observed and modeled distance modulus values $\mu_\text{obs} - \mu_\text{model}$ as a function of redshift.}
\label{fig:residuals}
\end{figure}

Figure~\ref{fig:mu_comparison} illustrates the comparison between the best-fit distance modulus curves \(\mu(z)\) for the spinor field Generalized Chaplygin Gas (GCG) model and the standard \(\Lambda\)CDM model, along with the binned Pantheon SN (SNb) data. Both models exhibit a good match with observational data across the redshift range \(0 < z \lesssim 1.6\), with the GCG curve almost indistinguishable from \(\Lambda\)CDM at most redshifts. To further quantify the fit, Figure~\ref{fig:residuals} shows the residuals \((\mu_{\text{obs}} - \mu_{\text{theory}})\) for the GCG model, where most points lie within a narrow band around zero, indicating the model's consistency with data. The small deviations at very low redshift due to local inhomogeneities,
peculiar velocities,  and high redshift bins are well within statistical uncertainties and reflect typical fluctuations due to data dispersion.

\begin{figure}[!ht]
\centering
\includegraphics[width=0.8\textwidth]{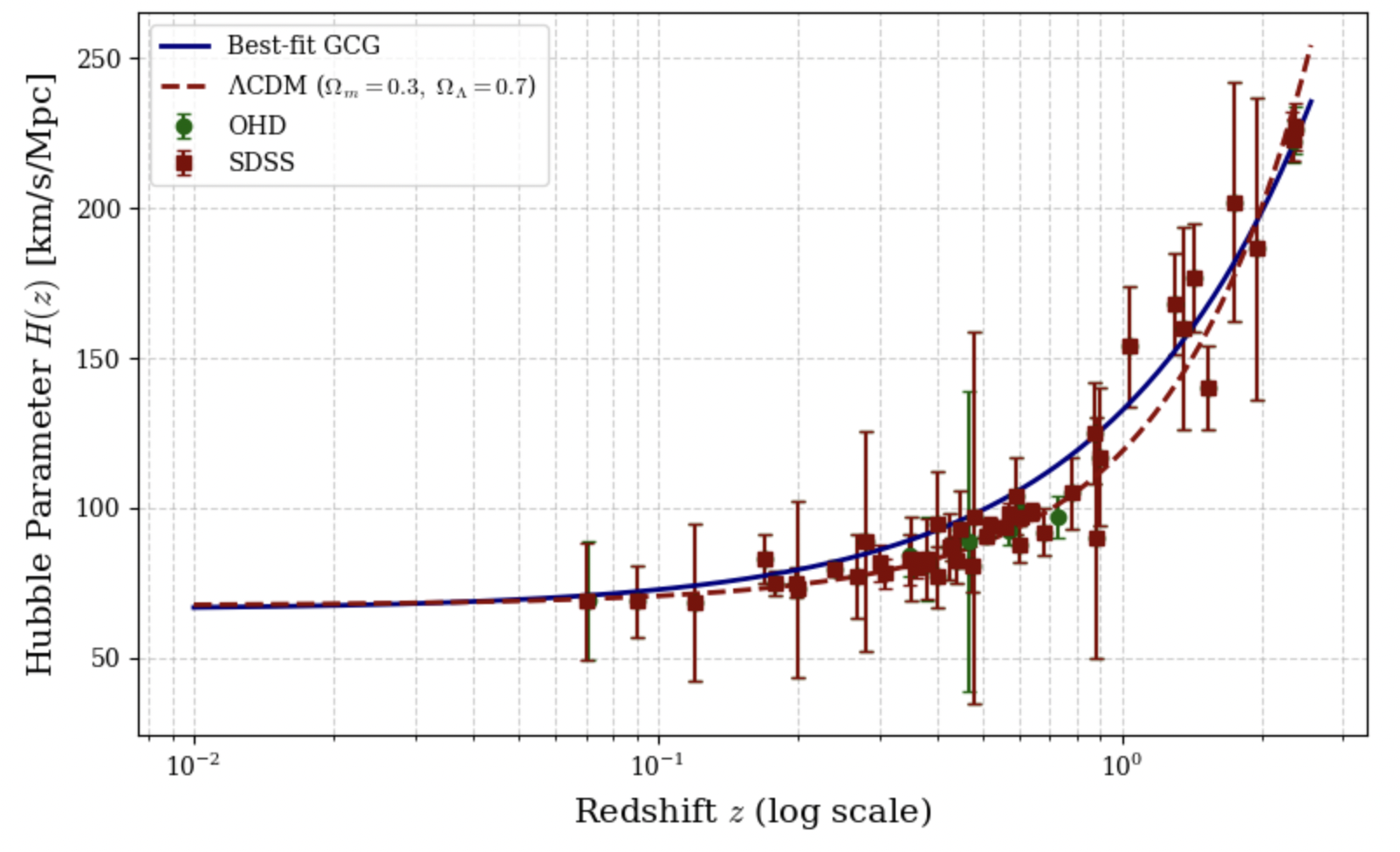}
\caption{Comparison of best-fit $H(z)$ of the spinor field GCG model to the Hubble data sets ($OHD+SDSS$) along with $\Lambda CDM$ model.}
\label{fig:Hubble_fit_comparison}
\end{figure}

Figure~\ref{fig:Hubble_fit_comparison} presents the evolution of the Hubble parameter \(H(z)\) as a function of redshift \(z\) in logarithmic scale, comparing the predictions of the best-fit spinor field (GCG) model with those of the standard \(\Lambda\)CDM model. The figure clearly shows that the GCG model provides an excellent fit to the data across the full redshift range, including the high-\(z\) regime where some deviation from the \(\Lambda\)CDM trajectory is noticeable. 

\begin{table}[htbp]
\centering
\caption{Comparison of statistical indicators between the best-fit Spinor GCG and $\Lambda$CDM models using the combined dataset (OHD + SDSS + SNb + BAO).}
\begin{tabular}{lccccccc}
\hline
\textbf{Model} & Parameters & \(\chi^2_{\mathrm{tot}}\) & \(\chi^2_{\nu}\) & AIC & BIC & \(\Delta\)AIC & \(\Delta\)BIC \\
\hline
\textbf{Spinor GCG}     & 3 & 134.742 & 1.061 & 140.742 & 149.345 & 0.000 & 0.000 \\
\textbf{$\Lambda$CDM}   & 2 & 139.191 & 1.079 & 143.191 & 148.941 & 2.449 & -0.404 \\
\hline
\end{tabular}
\label{tab:model_comparison}
\end{table}

To evaluate the statistical performance of the spinor GCG model in comparison with the standard $\Lambda$CDM cosmology, we analyze their fits to a combined observational dataset comprising OHD, SDSS, SNb, and BAO measurements. For the $\Lambda$CDM model, we perform MCMC sampling using this combined dataset and obtain the best-fit parameters: \(\Omega_m = 0.3000\) and \(H_0 = 67.00\) km\,s\(^{-1}\)\,Mpc\(^{-1}\). A summary of the model comparison is presented in Table~\ref{tab:model_comparison}, including the number of free parameters, total chi-squared \((\chi^2_{\mathrm{tot}})\), reduced chi-squared \((\chi^2_{\nu})\), Akaike Information Criterion (AIC), Bayesian Information Criterion (BIC), and the differences in AIC and BIC relative to the best-performing model.

\begin{figure}[htbp]
    \centering
    \begin{subfigure}{0.48\textwidth}
        \centering
        \includegraphics[width=\linewidth]{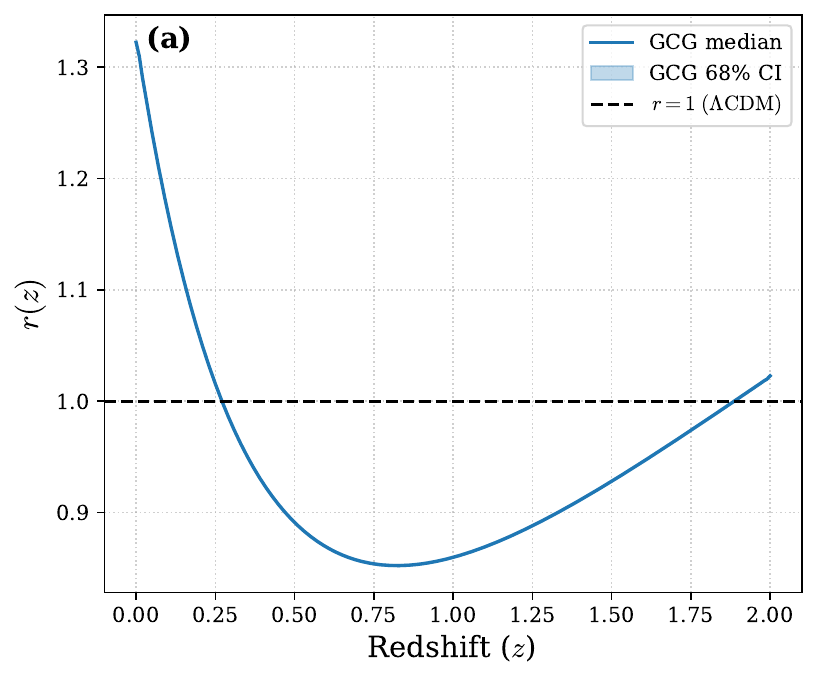}
    \end{subfigure}
    \hfill
    \begin{subfigure}{0.5\textwidth}
        \centering
        \includegraphics[width=\linewidth]{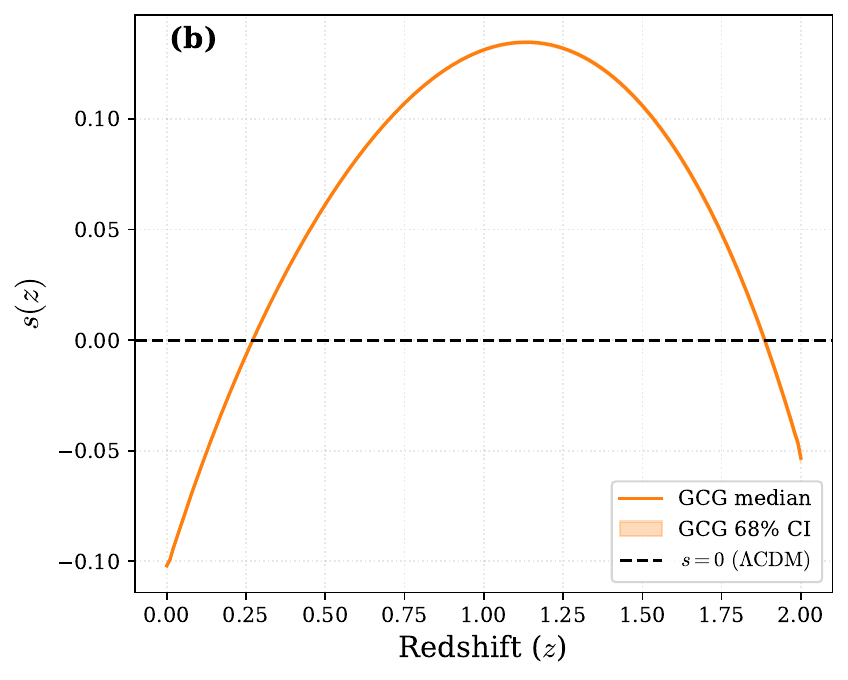}
    \end{subfigure}

    \vspace{-0.1em}
    \begin{subfigure}{0.55\textwidth}
        \centering
        \includegraphics[width=\linewidth]{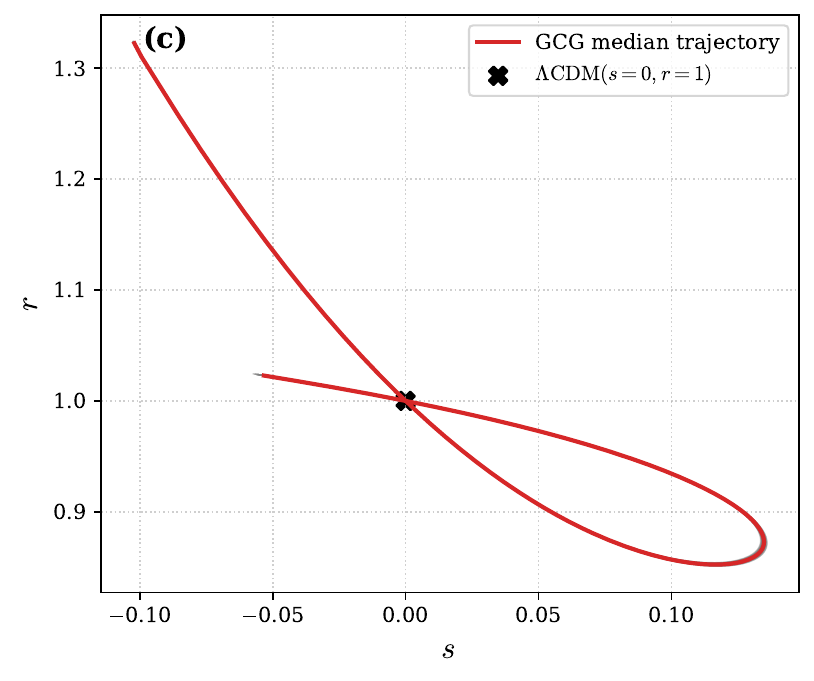}
    \end{subfigure}
    
    \caption{Statefinder diagnostics for the best-fit Generalized Chaplygin Gas (GCG) model using the combined dataset (OHD + SDSS + SNb + BAO). 
    (a) Evolution of the parameter $r(z)$. 
    (b) Evolution of the parameter $s(z)$. 
    (c) The $r$--$s$ plane trajectory, with the $\Lambda$CDM fixed point $(s=0,\,r=1)$ marked by a black cross. 
    Shaded regions denote the 68\% confidence intervals from the MCMC posterior distribution.}
    \label{fig:statefinder_all}
\end{figure}

 In terms of model selection criteria, the GCG model has the lowest Akaike Information Criterion (AIC = 140.742) and Bayesian Information Criterion (BIC = 149.345), with a reduced chi-squared \(\chi^2_{\nu} = 1.061\), while the corresponding values for $\Lambda$CDM are AIC = 143.191 and BIC = 148.941, \(\chi^2_{\nu} = 1.079\). The AIC difference of \(\Delta\mathrm{AIC} = 2.449\) suggests mild evidence in favor of the GCG model, while the marginally lower BIC value for $\Lambda$CDM (\(\Delta\mathrm{BIC} = -0.404\)) indicates that the simpler model may still be competitive under stronger penalization for model complexity. Overall, the spinor GCG model exhibits a statistically comparable and slightly improved fit to the observational data, reinforcing its potential as a viable alternative to the standard cosmological model in the late-time universe.

Additionally, at the present epoch we obtain the statefinder values $r_0 = 1.322$ and $s_0 = -0.102$, which are significantly offset from the $\Lambda$CDM fixed point $(r,s) = (1,0)$.
Figure~\ref{fig:statefinder_all} shows that the trajectory in the $r$--$s$ plane allows the GCG model to interpolate smoothly between the standard matter-dominated regime and a dark energy–dominated phase at late times, while retaining distinguishable features from $\Lambda$CDM at $z \simeq 0$.
This deviation highlights the model’s dynamical nature and its capability to account for late-time cosmic acceleration within a unified framework.

\section*{V. Conclusion}

In this work, we have proposed and analyzed a cosmological model where the Generalized Chaplygin Gas (GCG), representing a unification of dark energy and dark matter, arises from a nonlinear spinor field under spherically symmetric Friedmann–Lemaître–Robertson–Walker (FLRW) spacetime. The model is theoretically motivated as a means to explore the interplay between space-time geometry and fermionic fields, offering an alternative perspective on cosmic acceleration.

We performed a comprehensive statistical analysis by fitting the model to current observational datasets, including cosmic chronometers (OHD), SDSS data, baryon acoustic oscillations (BAO), and binned Pantheon Type Ia supernovae data. Using Markov Chain Monte Carlo (MCMC) methods, we obtained the best-fit parameters and compared the model’s performance against the standard $\Lambda$CDM model. Our results show that the spinor field GCG model provides a competitive fit to the data, with a slightly lower total $\chi^2$, reduced $\chi^2$, and AIC values compared to $\Lambda$CDM. Although the Bayesian Information Criterion (BIC) marginally favors the simpler $\Lambda$CDM due to its lower parameter count, the GCG model remains statistically viable. Importantly, the model yields a lower present-day Hubble constant $H_0$, potentially alleviating the Hubble tension between early and late-time measurements.

The redshift evolution of the equation of state $w(z)$ and the deceleration parameter $q(z)$ further supports the dynamical nature of the spinor GCG scenario. This dynamical behavior, rooted in geometry–spinor field interaction, offers a deeper physical insight into the nature of dark energy. The statefinder analysis further reveals present-day values of $r_0 = 1.322$ and $s_0 = -0.102$, marking a clear deviation from the $\Lambda$CDM fixed point $(1,0)$ and reflecting the GCG model’s dynamical nature in describing late-time cosmic acceleration.

Compared to other formulations of the Generalized Chaplygin Gas (GCG) model, typically introduced through phenomenological fluid descriptions, our spinor field GCG model introduces a deeper theoretical framework by embedding the GCG dynamics within a spinor field under a spherically symmetric FLRW spacetime. This geometric-spinor coupling adds a layer of fundamental physics, grounding the unification of dark matter and dark energy in quantum field dynamics rather than ad hoc prescriptions. Furthermore, unlike the standard \(\Lambda\)CDM model, which assumes a cosmological constant with a fixed equation of state, our model allows for a dynamic evolution of \(w(z)\), allowing for a more natural transition from matter-dominated to accelerated expansion phases.

In conclusion, the spinor field GCG model serves as a theoretically consistent and observationally supported framework that may provide an alternative to the $\Lambda$CDM paradigm. Future studies, including CMB constraints, structure formation, and nonlinear perturbation analysis, could further test the robustness and implications of this model in cosmology.
\\
\\
\\
\\
{\bf Declarations}

\vskip 5 mm {\bf Competing interests:} { There is
no conflict of interests.}


\vskip 5 mm {\bf Funding:}  Not applicable.

\vskip 5 mm {\bf Availability of data and materials:} No new data sets were generated during the current study.


\begin{thebibliography}{9999}
\bibitem{Planck2018}
{\it N. Aghanim et al. (Planck Collaboration)}, Astron. Astrophys. {\bf 641}, A6 (2020)

\bibitem{BOSS2017}
{\it S. Alam et al. (BOSS Collaboration)}, Mon. Not. R. Astron. Soc. {\bf 470}, 2617 (2017)

\bibitem{Pantheon2018}
{\it D. M. Scolnic et al.}, The Astrophysical Journal, {\bf 859}, 101 (2018)

\bibitem{PeeblesRatra2003}
{\it P. J. E. Peebles and B. Ratra}, Rev. Mod. Phys. {\bf 75}, 559 (2003)

\bibitem{Frieman2008}
{\it J. A. Frieman, M. S. Turner, and D. Huterer}, Ann. Rev. Astron. Astrophys. {\bf 46}, 385 (2008)

\bibitem{Schwarz2016}
{\it D. J. Schwarz, C. J. Copi, D. Huterer, and G. D. Starkman}, Class. Quantum Grav. {\bf 33}, 184001 (2016)

\bibitem{Riess2021}
{\it A. G. Riess et al.}, Astrophys. J. Lett. {\bf 908}, L6 (2021)

\bibitem{DiValentino2021}
{\it E. Di Valentino et al.}, Class. Quantum Grav. {\bf 38}, 153001 (2021)

\bibitem{Caldwell1998}
{\it R. R. Caldwell, R. Dave, and P. J. Steinhardt}, Phys. Rev. Lett. {\bf 80}, 1582 (1998)

\bibitem{ArmendarizPicon2001}
{\it C. Armendariz-Picon, V. Mukhanov, and P. J. Steinhardt}, Phys. Rev. D {\bf 63}, 103510 (2001)

\bibitem{Clifton2012}
{\it T. Clifton, P. G. Ferreira, A. Padilla, and C. Skordis}, Phys. Rep. {\bf 513}, 1 (2012)

\bibitem{Bertone2005}
{\it G. Bertone, D. Hooper, and J. Silk}, Phys. Rep. {\bf 405}, 279 (2005)

\bibitem{Kamenshchik2001}
{\it A. Y. Kamenshchik, U. Moschella, and V. Pasquier}, Phys. Lett. B {\bf 511}, 265 (2001)

\bibitem{Bento2002}
{\it M. C. Bento, O. Bertolami, and A. A. Sen}, Phys. Rev. D {\bf 66}, 043507 (2002)

\bibitem{Sandvik2004}
{\it H. B. Sandvik, M. Tegmark, M. Zaldarriaga, and I. Waga}, Phys. Rev. D {\bf 69}, 123524 (2004)

\bibitem{Ribas2005}
{\it M. O. Ribas, F. P. Devecchi, and G. M. Kremer}, Phys. Rev. D {\bf 72}, 123502 (2005)

\bibitem{Saha2006}
{\it B. Saha}, Phys. Rev. D {\bf 74}, 124030 (2006)

\bibitem{ArmendarizPicon2003}
{\it C. Armendariz-Picon and P. B. Greene}, Gen. Relativ. Gravit. {\bf 35}, 1637 (2003)

\bibitem{Kremer2003}
{\it G. M. Kremer}, Phys. Rev. D {\bf 68}, 123507 (2003)


\bibitem{Moresco2012}
{\it M. Moresco et al.}, JCAP {\bf 08}, 006 (2012)

\bibitem{Sahni2014}
{\it V. Sahni, A. Shafieloo, and A. A. Starobinsky}, Astrophys. J. {\bf 793}, L40 (2014)


\bibitem{Beutler2011}
{\it F. Beutler et al.}, Mon. Not. R. Astron. Soc. {\bf 416}, 3017 (2011)

\bibitem{Ross2015}
{\it A. J. Ross et al.}, Mon. Not. R. Astron. Soc. {\bf 449}, 835 (2015)

\bibitem{ForemanMackey2013}
{\it D. Foreman-Mackey, D. W. Hogg, D. Lang, and J. Goodman}, Publ. Astron. Soc. Pac. {\bf 125}, 306 (2013)

\bibitem{Narlikar} {\it Narlikar J.V.} Introduction to Relativity.
(Cambridge University Press, NY, 2010)

\bibitem{SahaPRD2001} {\it Saha B.}
Phys. Rev. D {\bf 64}, 123501 (2001).

\bibitem{bsaha2025} {\it Saha B.} Zurnal eksperimentalnoj i teoreticeskoj fiziki {\bf 167}(1) 49-58 (2025)



\bibitem{Moresco2016}
{\it M. Moresco et al.}, JCAP {\bf 05}, 014 (2016)
\bibitem{Alam2021}
{\it S. Alam et al.}, Phys. Rev. D {\bf 103}, 083533 (2021)

\bibitem{Bautista2021}
{\it J. E. Bautista et al.}, Mon. Not. R. Astron. Soc. {\bf 500}, 736 (2021)
\bibitem{6dFBAO2011}
{\it F. Beutler et al.}, Mon. Not. R. Astron. Soc. {\bf 416}, 3017 (2011)

\bibitem{SDSSMGS2015}
{\it A. J. Ross et al.}, Mon. Not. R. Astron. Soc. {\bf 449}, 835 (2015)

\bibitem{MCMC2002}
{\it A. Lewis and S. Bridle}, Phys. Rev. D, {\bf 66}, 103511 (2002).

\bibitem{Lewis2002_CPL_q0}
A. Lewis and S. Bridle, Phys. Rev. D {\bf 66}, 103511 (2002).

\bibitem{VargasDosSantos2015_q0}
M. Vargas dos Santos, R. R. R. Reis, and I. Waga, JCAP {\bf 05}, 014 (2015).

\bibitem{Santos2016_CPL}
{\it B. Santos, J. C. Carvalho, and J. S. Alcaniz}, 
Astropart. Phys. {\bf 79}, 40 (2016).


















\end{thebibliography}
\end{document}